\documentclass[vecphys]{svmult}

\usepackage{graphicx}        % standard LaTeX graphics tool
\usepackage[bottom]{footmisc}% places footnotes at page bottom

\newcommand{\Msun}{{M}_{\odot}}

\newcommand{\iso}[2]{\hbox{${}^{#1}{\rm #2}$}}

\begin{document}

\title*{Yields from single AGB stars}
\author{Amanda I. Karakas\inst{1}}
\institute{Research School of Astronomy \& Astrophysics, Mt Stromlo Observatory,
Cotter Road, Weston Creek, ACT 2611, Australia
\texttt{akarakas@mso.anu.edu.au}}
\maketitle

\begin{abstract}
In order to understand the composition of planetary nebulae we 
first need to study the nucleosynthesis occurring in the progenitor star
during the thermally-pulsing Asymptotic Giant Branch (AGB) phase.
I present an overview of single AGB evolution,
with an emphasis on the mixing processes that alter the envelope 
composition, followed by a discussion of the stellar yields 
available from single AGB stellar models.
\end{abstract}

\section{Introduction}
\label{sec:1}

The last nuclear burning phase of a low to intermediate-mass 
($\sim 0.8$ to 8$\Msun$) star's life is the thermally-pulsing asymptotic
giant branch (TP-AGB). The outer envelope is lost by low velocity 
stellar winds during the AGB, with the termination being the final ejection
of the envelope. The star then evolves through the brief post-AGB and 
planetary nebula phases before ending its life as a white dwarf. 
The gaseous nebula is the remnant of the envelope that once 
surrounded the core, that is now exposed as the central star 
of the illuminated nebula. The abundances of the nebula can reveal
information about stellar nucleosynthesis and mixing
during the AGB. For this reason PN abundances could, in principle, 
be used to constrain stellar models.

In this proceedings I summarize the evolution and nucleosynthesis
during the AGB.  I briefly discuss the differences 
between ``synthetic'' and ``detailed'' AGB models,
and compare yields from different authors. 

\section{Asymptotic giant branch stars} \label{sec:agb}

The structure and evolution of low and intermediate mass stars 
prior to and during the AGB has been previously discussed by
\cite{busso99,herwig05}; see \cite{vanwinckel03} for a review 
of post-AGB stars. All stars begin their nuclear-burning life 
on the main sequence. Following core H exhaustion the core
contracts, the outer layers expand and the star becomes
a red giant, characterized by an inert He core, an H-burning 
shell, and a deep convective envelope that extends to the 
stellar surface. It is during the ascent of the giant branch 
that the inner edge of the convective envelope moves inward in 
mass and the first dredge-up (FDU) occurs, where partially 
H-processed material (e.g. \iso{4}He, \iso{13}C, \iso{14}N) 
is mixed to the surface.

After a phase of central He-burning, the core contracts and 
there is a structural re-adjustment to shell He burning. The 
re-adjustment drives a strong expansion of the outer layers 
and the star becomes a red giant for the second time; the 
star is now said to be on the AGB.  For stars with $m \ge 4\Msun$ 
the convective envelope moves inward to regions where complete
H burning (mostly \iso{4}He and \iso{14}N) had previously 
occurred, this is the second dredge up (SDU). 

\begin{figure}
\begin{center}
\includegraphics[scale=0.35, angle=0]{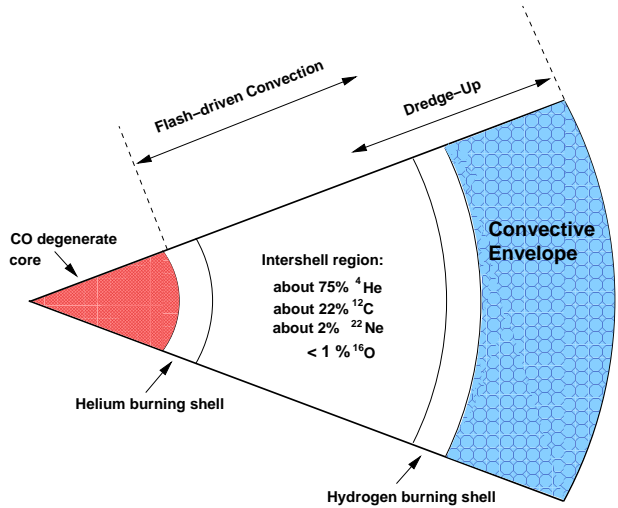}\\
\caption{Schematic structure of an AGB star.
\label{agb}}
\end{center}
\end{figure}

An AGB star is characterized (Fig.~\ref{agb}) by two nuclear 
burning shells, one burning He above a degenerate C-O core and
another burning H, below a deep convective envelope.
In between lies the intershell region composed mostly of \iso{4}He.
The He-burning shell is thermally unstable, flashing 
every 10$^{4}$~years or so. The energy produced by the thermal pulse
(TP) drives a convective pocket in the He-intershell which
acts to homogenize abundances within that region.
After the occurrence of a TP the convective envelope may move 
inwards and mix products of partial He-burning 
(mostly \iso{4}He left unburnt and \iso{12}C, see 
Fig.~\ref{agb}) from the core
to the stellar surface. This is the third dredge-up (TDU), and 
is the mechanism responsible for turning (single) stars into 
C stars, where C/O $ > 1$ in the surface layers.  The TDU 
also mixes heavy elements produced by the $s$ process from 
the He-shell to the surface, where they were created during the 
previous interpulse \cite{gallino98}.  Following dredge-up, 
the star contracts, the H-shell is re-ignited and the star 
enters the interpulse phase where H-burning provides most 
of the luminosity.

Hot bottom burning (HBB) can occur for stars with $m \ge 4\Msun$,
when the base of the convective envelope dips into the top of 
the H-shell resulting in a thin layer hot enough to sustain 
proton-capture nucleosynthesis.  Observational
evidence for HBB includes the lack of bright C-rich
AGB stars in the  Large and Small Magellanic Clouds (LMC and SMC)
\cite{wood83}; many of these stars are also rich in Li and $s$ process
elements.  HBB converts \iso{12}C into \iso{14}N and
will prevent the atmosphere from becoming C rich
\cite{boothroyd93}. The copious amounts of \iso{14}N produced 
in this case will be primary\footnote{produced from the H and 
\iso{4}He initially present in the star.} owing to the primary 
\iso{12}C being dredged from the He-shell.  Intermediate-mass
evolution is sensitive to the initial composition and the mass-loss
law used in the calculation \cite{frost98a}, where the minimum stellar
mass for HBB is pushed to lower mass in lower metallicity models.

There is evidence from AGB stellar spectra \cite{abia97}, and from 
pre-solar grains \cite{nollett03} that ``extra-mixing'' 
processes are also operating in low-mass ($m \le 2\Msun$) AGB stars, 
along with the TDU.  The physical mechanism(s) responsible for the 
extra mixing are not known, although various processes have been 
proposed including rotation and thermohaline mixing.
The presence of a binary companion could have dramatic consequences
for the evolution and nucleosynthesis, however I leave discussions 
to Izzard, Taam and Podsiadlowski (these proceedings).

\section{Synthetic versus detailed AGB models}
\label{sec:synth}

Owing to fact that calculating a TP-AGB model is a computationally 
intensive task, synthetic AGB models, which use fitting formulae 
to model the evolution quickly, have proved to be a successful approach 
for population syntheses studies that require $N \sim 10^{6}$ stars. 
Historically this approach was validated by the fact that the stellar 
luminosity on the AGB is nearly a linear function of the core mass 
\cite{paczynski75,groen93}, although this relation breaks down for 
stars undergoing HBB \cite{bloecker91}. Synthetic AGB models have 
successfully been used to model AGB populations \cite{groen93}, 
and compute stellar yields \cite{vandenhoek97,marigo01,izzard04b}.

Many of the parameterizations used in synthetic evolution 
studies are derived from detailed stellar models, such as the 
growth of the H-exhausted core with time, and as such are only 
accurate over the range in mass and metallicity of the stellar models 
they are based upon. An example is provided by \cite{vandenhoek97} 
who compute AGB yields for initial masses between 0.9 and 8$\Msun$ 
whereas the interpulse-period-core mass relation \cite{boothroyd88c}
they use was only derived for stars with initial masses between 1 
and 3$\Msun$.  What affect this has on the yields is unclear since
this relation will affect the number of TPs during the TP-AGB phase 
and hence the level of chemical enrichment. Recent improvements in 
computer power mean that  grids of detailed AGB models can now be 
produced in a reasonable time \cite{karakas03b,karakas06a,herwig04b};
however producing yields from $N \ge 20$ AGB stars for any given 
metallicity range is still challenging. For this reason synthetic 
models are still preferred for some applications.

\begin{figure}
\begin{center}
\includegraphics[scale=0.37, angle=270]{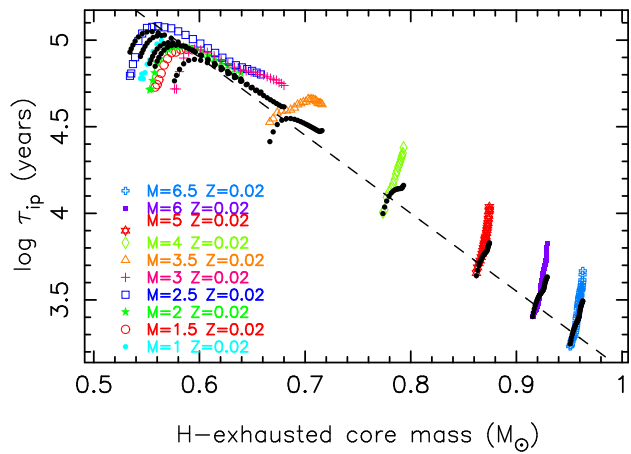}\\
\caption{Interpulse period ($\log \tau_{\rm ip}$ years) as a 
function of core mass ($\Msun$) for the $Z=0.02$ 
stellar models (colored dots). The fits from 
\cite{boothroyd88c} (dashed line) and \cite{wagenhuber98}
(solid black dots) are shown for comparison.
\label{tau-ip}}
\end{center}
\end{figure}

An example of the difference between detailed models and the fits
used in synthetic AGB algorithms is shown in Fig.~\ref{tau-ip} 
for the interpulse-period core mass relation.  We show results
from detailed AGB models of solar composition against two commonly 
used fits. The fits are a reasonable match to the detailed models 
with small core masses ($M_{\rm c} \le 0.7\Msun$), but 
under-estimate the growth of the interpulse period for
larger core masses.

\section{The stellar yields from single AGB models}
\label{sec:yields}

\begin{figure}
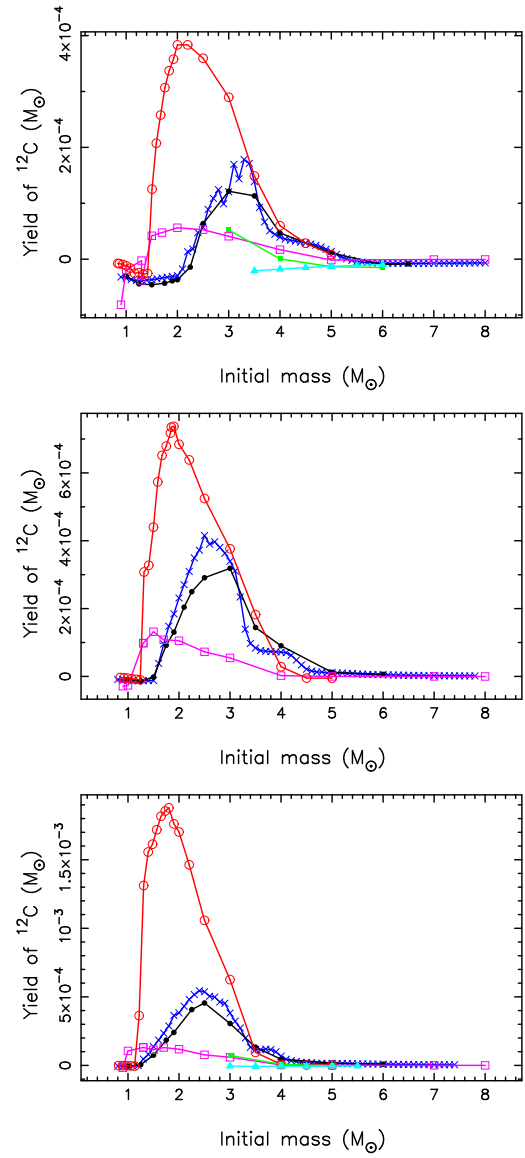

\begin{center}
\begin{tabular}{c}
\includegraphics[scale=0.30, angle=270]{fig3a.ps}\\
\includegraphics[scale=0.30, angle=270]{fig3b.ps}\\
\includegraphics[scale=0.30, angle=270]{fig3c.ps}
\end{tabular}
\caption{Weighted yield of \iso{12}C as a function of the initial
mass for the $Z=0.02$ (top), the $Z=0.008$ (middle) and the 
$Z=0.004$ models (bottom). We show results from
our calculations (black solid points), \cite{vandenhoek97} (open
magenta squares), \cite{forestini97} (solid green squares), \cite{marigo01}
(open red circles), \cite{ventura02} (solid aqua triangles) and
\cite{izzard04b} (blue crosses). \cite{forestini97} and \cite{ventura02}
do not provide yields for $Z=0.008$ and cover a
narrower mass range, between 2.5 and 6$\Msun$.
}\label{c12}
\end{center}
\end{figure}

In \cite{karakas07b} we present results from grids of AGB
models, including data about the stellar structure and the yields.
These are also available for download from:
{\tt http://www.mso.anu.edu.au/\~{}akarakas/model\_data/} and \\
{\tt http://www.mso.anu.edu.au/\~{}akarakas/stellar\_yields/}.
~More details about the numerical technique and the yields can 
be found in \cite{karakas07b}, but we note here that yield can 
be negative, in the case where the element is destroyed, 
and positive if it is produced.

In Fig.~\ref{c12} we show the AGB yields
of \iso{12}C as a function of the initial stellar mass.
The yields have been weighted by the initial mass function 
(IMF) of \cite{kroupa93}, and we show results from the 
$Z=0.02$ (solar), 0.008 (LMC) and 0.004 (SMC) metallicity 
models. For comparison we also show the yields from a number 
of different synthetic AGB calculations, and from \cite{ventura02}; 
see the figure caption for details.  The yield of \iso{12}C 
is representative of low-mass AGB nucleosynthesis where the 
main contributor to the yield is either the FDU at 
low mass ($m \le 1.2\Msun$), where the \iso{12}C surface 
abundance decreases, or from the TDU where we find 
substantial increases. Models with $m > 4\Msun$ produce 
little \iso{12}C at these metallicities owing to HBB. 
The yields of \iso{14}N instead peak in this mass range, 
even after weighting the yields with an IMF \cite{karakas07b}.
HBB may also produce \iso{23}Na, \iso{26}Al and 
the heavy Mg isotopes through the combined operation 
of the TDU and the MgAl chains \cite{karakas03b,karakas06a}.

Our yields are similar in behaviour to those of \cite{marigo01}, 
although because her models have deeper TDU at a lower core mass, 
the yields of He-shell material e.g. \iso{4}He, \iso{12}C are 
higher. We notice significant differences with \cite{vandenhoek97} 
especially in regards to \iso{14}N. This is owing to their simplistic 
treatment of HBB nucleosynthesis which under-predicts the amount 
of CNO cycling compared to all other computations. The yields 
from \cite{vandenhoek97} for masses between $\sim$6 to 8$\Msun$ 
are based on extrapolations of fitting formula to this mass range
and should be treated with caution. Their models also 
produce much more \iso{16}O at $m \sim 1\Msun$ 
at $Z=0.02$; this is especially noticeable when weighting 
by the IMF. This was also found by \cite{izzard04b}
and the reasons for the production are unclear because 
the FDU should not enhance the surface in \iso{16}O.

\begin{figure} 
\begin{center}
\begin{tabular}{c}
\includegraphics[scale=0.4, angle=270]{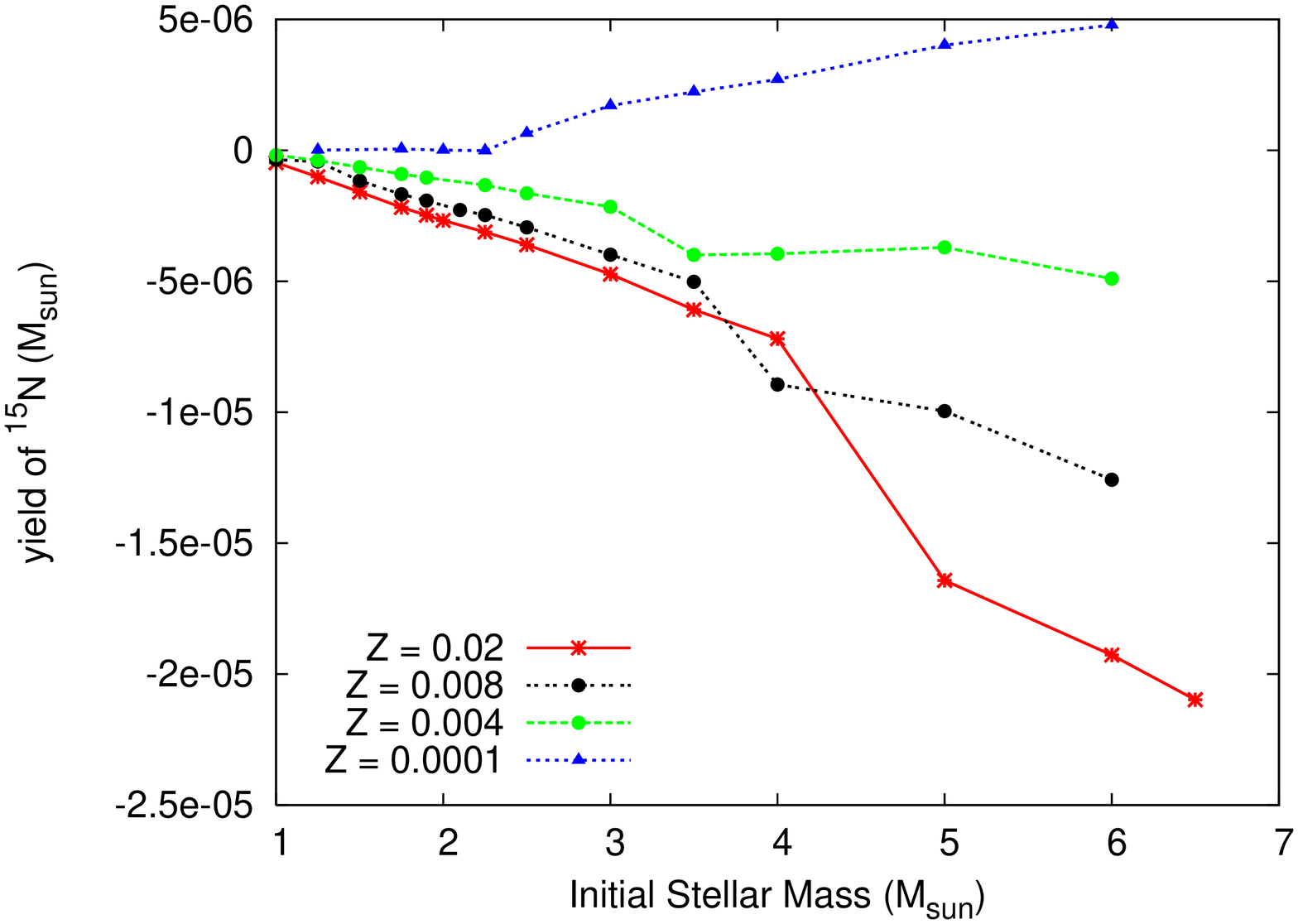} \\
\includegraphics[scale=0.4, angle=270]{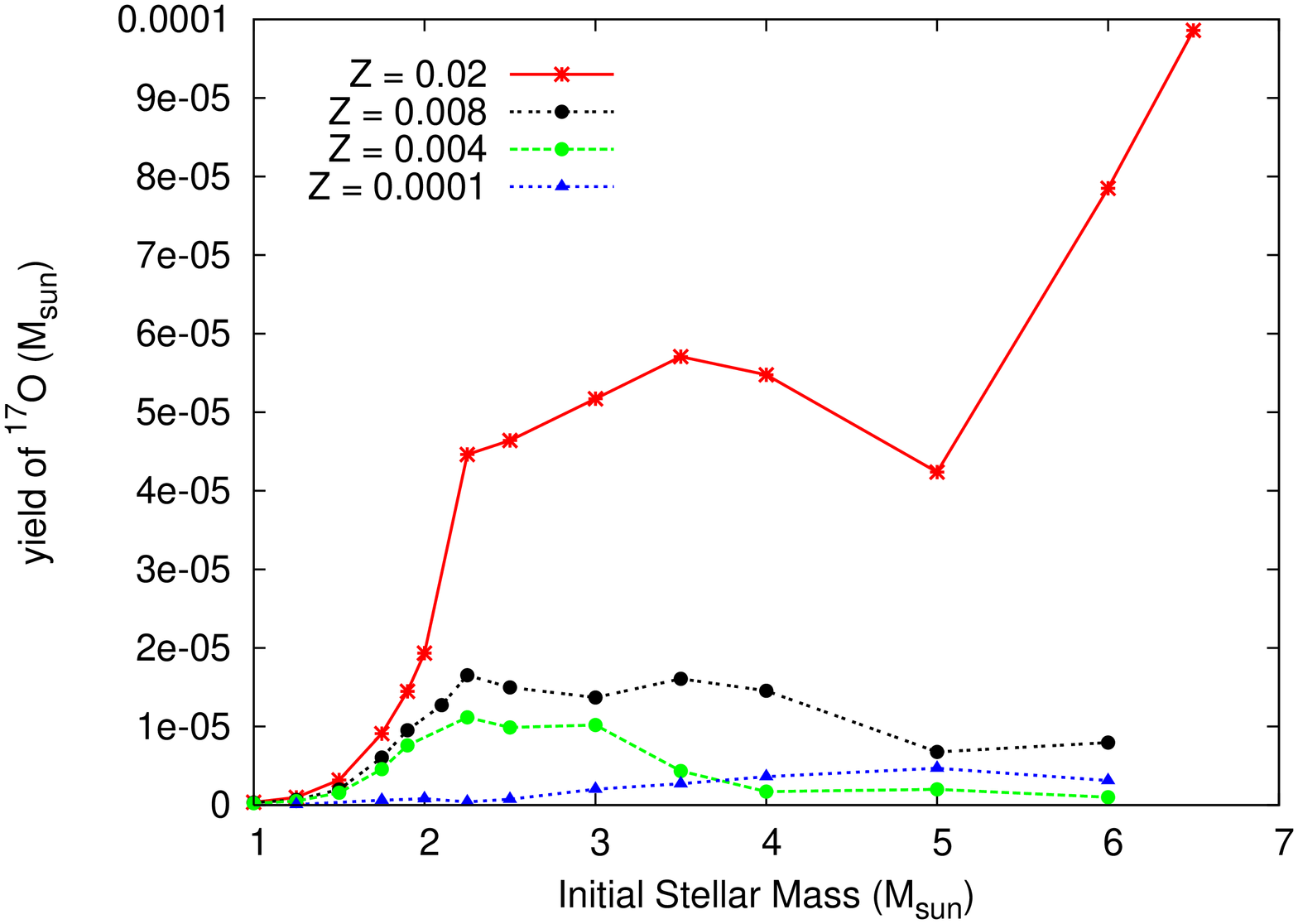}\\
\end{tabular}
\caption{Yield of \iso{15}N (top panel) and \iso{17}O 
(lower panel) as a function of the initial mass and
metallicity.
\label{minor}}
\end{center}
\end{figure}

In Fig.~\ref{minor} we show the yields of \iso{15}N and 
\iso{17}O as a function of initial mass and metallicity. 
The unexpected result is the production of \iso{15}N in the
lowest $Z$ AGB models; this is owing to leakage from the
\iso{14}N(p,$\gamma$)\iso{15}O reaction at very high temperature. 
For the \iso{15}N yields from the more metal-rich models, and 
for \iso{17}O, the yields scale with $Z$ as expected.
Fig.~\ref{minor} shows that low-mass AGB stars can 
equally participate in the production of \iso{17}O,
whereas this has been previously ignored in 
chemical evolution studies e.g. \cite{romano03} probably 
owing to a lack of available yields.

\subsection{Yields for planetary nebulae}

\begin{table}
\begin{center}
\caption{PN yields 3$\Msun$, $Z = 0.02$ model.}\label{pnyield}
\begin{tabular}{cccccc}
\hline Isotope &  $A$ & yield & $X0(i)$  &  $\langle X(i) \rangle$ & $N(i)/N(H)$ \\
\hline
\hline
\iso{4}He   &   4 &  4.39725E-02 &  2.92881E-01 & 3.21630E-01 & 1.23369E-01 \\
\iso{12}C   &  12 &  1.10031E-02 &  3.40894E-03 & 1.06028E-02 & 1.35560E-03 \\
\iso{13}C   &  13 &  7.72216E-05 &  4.10914E-05 & 9.15790E-05 & 1.08084E-05 \\
\iso{14}N   &  14 &  2.23653E-03 &  1.05449E-03 & 2.51674E-03 & 2.75816E-04 \\
\iso{15}N   &  15 & $-$3.12235E-06 &  4.14117E-06 & 2.09975E-06 & 2.14776E-07 \\
\iso{16}O   &  16 & $-$8.21213E-04 &  9.60272E-03 & 9.06575E-03 & 8.69348E-04 \\
\iso{17}O   &  17 &  3.40679E-05 &  3.87707E-06 & 2.61507E-05 & 2.36017E-06 \\
\iso{18}O   &  18 & $-$8.87124E-06 &  2.16060E-05 & 1.58060E-05 & 1.34728E-06 \\
\hline
\hline
\end{tabular}
\end{center}
\end{table}

In \cite{karakas07b} we present yields for PN, where we
integrate the amount of matter lost over the last two TPs. 
In Table~\ref{pnyield} we show an example of the yields 
for a 3$\Msun$ progenitor of solar composition. 
The columns contain the species $i$, the atomic weight $A(i)$, 
followed by the net yield (in $\Msun$; see \cite{karakas07b} 
for the definition), the average mass fraction of $i$ lost in 
the wind from the last two TPs, $\langle X(i) \rangle$,
and the initial mass fraction $X0(i)$. 
The final column is the abundance of $i$ by number compared 
to the number of hydrogen atoms, $N_{i}/N_{\rm H}$,
in the matter lost over the final two TPs.
The online tables include yields for all 74 species included
in the nuclear network. 

These yields can be directly compared to the composition
of PNe and as such could be useful for comparisons to objects
in the Galaxy, LMC and SMC. This is because the initial C, N 
and O abundances used in the stellar models were taken from
abundances derived from HII regions of the Magellanic
Clouds, see \cite{karakas07b} for details.

%
% Use the following syntax and markup for your references
%

%%%%%%%%%%%%%%%%%%%%%%%%%%%%%%%%%%%%%%%%%%%%%%%%%%%%%%%%%%%%%%%%%%%%%%  }

%%%%%%%%%%%%%%%%%%%%%%%%%%%%%%%%%%%%%%%%%%%%%%%%%%%%%%%%%%%%%%%%%%%%%%


\begin{thebibliography}{99.}
\bibitem{abia97}{Abia}, C. \& {Isern}, J. 1997, MNRAS, 289, L11

\bibitem{bloecker91}
{Bloecker}, T. \& {Schoenberner}, D. 1991, A\&A, 244, L43

\bibitem{boothroyd88c}
{Boothroyd}, A.~I. \& {Sackmann}, I.-J. 1988, ApJ, 328, 653

\bibitem{boothroyd93}
{Boothroyd}, A.~I., {Sackmann}, I.-J., \& {Ahern}, S.~C. 1993, ApJ, 416, 762

\bibitem{busso99}
{Busso}, M., {Gallino}, R., \& {Wasserburg}, G.~J. 1999, ARA\&A, 37, 239

\bibitem{forestini97}
{Forestini}, M. \& {Charbonnel}, C. 1997, A\&AS, 123, 241

\bibitem{frost98a} {Frost}, C.~A., et al. 1998, A\&A, 332, L17

\bibitem{gallino98}
{Gallino}, R., et al. 1998, ApJ, 497, 388

\bibitem{groen93}
{Groenewegen}, M.~A.~T. \& {de Jong}, T. 1993, A\&A, 267, 410

\bibitem{herwig04b}
 {Herwig}, F. 2004, ApJS, 155, 651

\bibitem{herwig05}
{Herwig}, F. 2005, ARA\&A, 43, 435

\bibitem{izzard04b}
{Izzard}, R.~G., et al. 2004, MNRAS, 350, 407

\bibitem{karakas03b}
{Karakas}, A.~I. \& {Lattanzio}, J.~C. 2003, PASA, 20, 279

\bibitem{karakas06a}
{Karakas}, A.~I., et al.  2006, ApJ, 643, 471

\bibitem{karakas07b}
{Karakas}, A.~I. \& {Lattanzio}, J.~C. 2007, PASA, submitted

\bibitem{kroupa93}
{Kroupa}, P., {Tout}, C.~A., \& {Gilmore}, G. 1993, MNRAS, 262, 545

\bibitem{marigo01}
{Marigo}, P. 2001, A\&A, 370, 194

\bibitem{nollett03}{Nollett}, K.~M., {Busso}, M., \& {Wasserburg}, G.~J. 2003, 
ApJ, 582, 1036

\bibitem{paczynski75}
{Paczynski}, B. 1975, ApJ, 202, 558

\bibitem{romano03}
Romano, D. \& Matteucci, F. 2003, MNRAS, 342, 185

\bibitem{wagenhuber98}
{Wagenhuber}, J. \& {Groenewegen}, M.~A.~T. 1998, A\&A, 340, 183

\bibitem{wood83}
{Wood}, P.~R., {Bessell}, M.~S., \& {Fox}, M.~W. 1983, ApJ, 272, 99

\bibitem{vandenhoek97}
{van den Hoek}, L.~B. \& {Groenewegen}, M.~A.~T. 1997, A\&AS, 123, 305

\bibitem{vanwinckel03}
{van Winckel}, H. 2003, ARA\&A, 41, 391

\bibitem{ventura02}
{Ventura}, P., {D'Antona}, F., \& {Mazzitelli}, I. 2002, A\&A, 393, 215

\end{thebibliography}
\end{document}